\documentstyle[aps,pra,epsf,floats]{revtex}
\def\figuresize{\ifpreprintsty 10cm \else 8cm \fi}

\begin{document}

\ifpreprintsty \else
\twocolumn[\hsize\textwidth\columnwidth\hsize\csname@twocolumnfalse%
\endcsname \fi

\draft
\title{Damage spreading in random field systems}

\author{Thomas Vojta}
\address{Institut f\"ur Physik, Technische Universit\"at, D-09107 Chemnitz, Germany and\\
                Materials Science Institute, University of Oregon, Eugene, OR97403, USA}
\date{version May 29, printed \today}
\maketitle

\begin{abstract}
We investigate how a quenched random field influences the damage spreading transition
in kinetic Ising models.  To this end we generalize a recent master equation approach 
and derive an effective field theory for damage spreading in random field systems.
This theory is applied  to the Glauber Ising model with a bimodal random field distribution.
We find that the random field influences the spreading transition by two different mechanisms
with opposite effects. First, the random field favors the same particular direction of 
the spin variable at each site in both systems which reduces the damage. Second,
the random field suppresses the magnetization which in turn tends to increase the damage.
The competition between these two effects leads to a rich behavior.
\end{abstract}

\ifpreprintsty \else
] \fi              


The central question of  damage spreading (DS)  \cite{kauffman,derrida1,stanley} 
is how a small perturbation in a cooperative system changes during the time evolution.
This is analogous to the question to what extent the time evolution depends on the 
initial conditions, one of the main questions in non-linear dynamics that lead to the
discovery of chaotic behavior \cite{schuster}.  
In order to study  DS  the simultaneous time evolution of two replicas 
of a cooperative system is considered. The two replicas
evolve stochastically under the same noise 
realization (i.e. the same random numbers are used in a  Monte-Carlo procedure).  
The differences in the microscopic configurations of the two replicas are then used to characterize 
the dynamics and to distinguish regular and chaotic phases, depending on external parameters.

Among the simplest cooperative systems are kinetic Ising models where DS
has been investigated quite intensively within the last years using different dynamical 
algorithms such as Glauber \cite{stanley,ts,grassberger,vojta1,vojta2} or heat-bath dynamics
\cite{derrida1,vojta1,grassheat,heat}.  In contrast to the equilibrium critical behavior the results
of DS do depend on the particular choice of the dynamical algorithm
although recently an attempt has been made to give a more objective definition of
DS \cite{hinrichsen}. In general, there are two different mechanisms by which
damage can spread in a kinetic Ising model. First, the damage can spread {\em within} a single
ergodic component (i.e. a pure state or free energy valley)
of the system. This is the case for Glauber or Metropolis dynamics. 
Second, the damage can spread when the system
selects one of the free energy valleys at random after a quench from high
temperatures to below the equilibrium critical temperature. 
This is the only mechanism to produce DS in an Ising
model with heat-bath dynamics. This algorithm is thus well suited for exploring
the structure of the free energy landscape.

In the literature the name DS has been applied not only
to the studies discussed above but also to a different
though related type of investigations in which the two systems are {\em not}
identical. Instead, one or several spins in 
one of the copies are permanently fixed in one direction. Thus the
equilibrium properties of the two replicas deviate from each other and 
their microscopic differences can be related to
equilibrium correlation functions \cite{coniglio,glotzer}. 
Note that in these  works the use of identical noise (i.e. random numbers) for the
two systems is not essential but only a method to reduce the statistical error.

Whereas DS in clean Ising models is comparatively well understood
less is known about disordered models. 
The influence of random fields has been investigated in a
two-dimensional Ising-like model with 
Metropolis dynamics giving a reduction of the damage at high temperatures 
but an increase at low temperatures \cite{wappler}.
By using the heat-bath algorithm
DS has been used to study the phase space structure of Ising spin 
glasses \cite{derrida2,campbell,tamarit} and the corresponding
critical behavior at the DS transition \cite{wang}.

In this Letter we consider the original DS problem, viz. the time evolution of
two identical systems and study the 
influence of a quenched random field on
DS in kinetic Ising models. To this end we
generalize the master equation approach \cite{vojta1,vojta2} to random field 
systems. The resulting effective field theory of DS is then applied to the Glauber
Ising model with a bimodal random field distribution. We study the dependence
of the spreading transition on temperature and field and determine the
phase diagram.

We consider two identical Ising models with $N$ sites described by
the Hamiltonians $H^{(1)}$ and $H^{(2)}$ given by
\begin{equation}
H^{(n)} = - \frac 1 2  \sum_{ij} J_{ij} S_i^{(n)} S_j^{(n)} - \sum_i \varphi_i S_i^{(n)} 
\end{equation}
where $S_i^{(n)}$ is an Ising variable with the values $\pm 1$ and $n=1,2$ 
distinguishes the two replicas. $J_{ij}$ is the (non-random)
exchange interaction between the spins. The random field values $\varphi_i$
are chosen independently from a distribution $\rho(\varphi)$. 
The dynamics of the systems are given by  stochastic maps
$S_i^{(n)}(t+1)= F[\{S_j^{(n)}(t)\}]$, e.g.,
the Glauber algorithm
\begin{equation}
S_i^{(n)} (t+1) = {\rm sgn} \left[ v[h_i^{(n)}(t)] 
- \frac 1 2 +S_i^{(n)}(t) \left( \xi_i(t) - \frac 1 2 \right) \right]
\end{equation}
where the transition probability $v(x)$ is given by the usual
Glauber expression
\begin{equation}
v(x) = {e^{x/T}/ ({e^{x/T}+ e^{-x/T}}}).
\end{equation}
Here $h_i^{(n)}(t)=\sum_j J_{ij} S_j^{(n)}(t) + \varphi_i$ 
is the local magnetic field at site $i$ and (discretized)
time $t$ in the system $n$. $\xi_i(t) \in [0,1)$ is a random number which is identical
for both systems, and $T$ denotes the temperature. 

Within the master equation approach \cite{vojta1,vojta2} the 
simultaneous time evolution of the two replicas is described
by the probability distribution
\begin{equation}
P(\nu_1,\ldots,\nu_N,t) = \left \langle \sum_{\nu_i(t)} \prod_i 
\delta_{\nu_i,\nu_i(t)} \right \rangle
\end{equation}
where $\langle \cdot \rangle$ denotes the average over 
the noise realizations. The variable $\nu_i$ with the values
$++, +-, -+$,  or $--$ describes the
states of the spin pair ($S_i^{(1)}, S_i^{(2)}$). The distribution 
$P$ fulfills the master equation
\begin{eqnarray}
\lefteqn{\frac d {dt} P(\nu_1,\ldots,\nu_N,t) =} \nonumber \\
& &- \sum_{i=1}^N \sum_{\mu_i \not= \nu_i}
P(\nu_1,\ldots,\nu_i,\ldots,\nu_N,t) w(\nu_i \to \mu_i) \nonumber\\
& &+ \sum_{i=1}^N \sum_{\mu_i \not= \nu_i}
P(\nu_1,\ldots,\mu_i,\ldots,\nu_N,t) w(\mu_i \to \nu_i). 
\label{eq:master}
\end{eqnarray}
The transition probabilities $w(\mu_i \to \nu_i)$ have to be calculated 
from the properties of the stochastic map $F$ which defines the dynamics.

As in the clean case we derive an effective field theory by assuming
that fluctuations at different sites are statistically independent which amounts
to approximating the distribution $P(\nu_1,\ldots,\nu_N,t)$ by a product 
of single-site distributions $P_{\nu_i}(i,t)$. However, in a disordered 
system different sites are not equivalent and thus their $P_{\nu_i}(i,t)$ are not identical.
This is the main difference to the clean case \cite{vojta1,vojta2} where
the single-site distributions are all identical. In the following we further assume 
that $P_{\nu_i}(i,t)$ is determined by the local value $\varphi_i$ of the random
field only, $P_{\nu_i}(i,t) \equiv P_{\nu_i}(\varphi_i,t)$.
In general, this is an approximation since sites with identical
random field values may well have different environments which should
influence the distribution. In the mean-field limit
of infinite dimensions oder infinite-range interactions, however, 
the above assumption becomes exact.

Inserting the decomposition
\begin{equation}
P(\nu_1,\ldots,\nu_N,t) = \prod_{i=1}^N P_{\nu_i}(\varphi_i,t).
\end{equation}
into the master equation (\ref{eq:master}) gives a system of coupled
equations of motion for the single-site distributions
\begin{eqnarray}
\frac d {dt} P_{\nu}(\varphi) =  \sum_{\mu \not= \nu} [&-& P_{\nu}(\varphi) W(\nu \to \mu,\varphi) 
\nonumber \\
&+& P_{\mu}(\varphi) W(\mu \to \nu,\varphi)],
\label{eq:ssmaster}
\end{eqnarray}
where $W(\mu \to \nu,\varphi)$ 
is the transition probability $w$ averaged over the states $\nu_i$ of all sites.
We now define the local damage 
$d(\varphi) = P_{+-}(\varphi) + P_{-+}(\varphi)$.
The total damage is obtained as the corresponding disorder average
\begin{equation}
D = \left \langle \frac 1 {2N} \sum_{i=1}^N |S_i^{(1)} - S_i^{(2)} |
\right \rangle = \int d\varphi
\, \rho(\varphi) \, d(\varphi) .
\end{equation}
The equation of motion of the local damage can be easily determined from 
(\ref{eq:ssmaster}). Using some symmetry relations \cite{vojta2} 
between the transition probabilities $W$, it reads
\begin{eqnarray}
\frac d {dt}&&d(\varphi)=
\label{eq:damaster} \\
&&=[1-d(\varphi)] [ W(-- \rightarrow +-,\varphi) + W(-- \rightarrow -+,\varphi)] 
\nonumber \\
&&+ d(\varphi) [-1 + W(-+ \rightarrow +-,\varphi) + W(-+ \rightarrow -+,\varphi)]
\nonumber
\end{eqnarray}

So far the considerations have been rather general, being valid for any 
(single-site) dynamic rule and any distribution of the random field. We now apply
the formalism to the Glauber Ising model in the mean-field limit $J_{ij} = J_0/N$
(for all $i$ and $j$). The random
field distribution remains unspecified so far. In order to determine the spreading 
point for infinitesimal initial damage it is sufficient to solve (\ref{eq:damaster}) in 
linear order in 
$d(\varphi)$. After calculating the transition probabilities $W$ in analogy
to the clean case \cite{vojta2} and some further algebra we obtain
\begin{equation}
\frac d {dt} d(\varphi) = - | m(\varphi) | \, d(\varphi) + \frac {J_0} T [1- m^2(\varphi)] \,D.
\label{eq:dalin}
\end{equation}
If we concentrate on DS processes starting in equilibrium conditions 
the local magnetization
\begin{equation}
m(\varphi) = \tanh [(J_0 m + \varphi)/T]
\label{eq:locmag}
\end{equation}
and the average magnetization $m$ are time-independent.
Eq. (\ref{eq:dalin}) is very similar to the corresponding eq.
(36) of Ref. \cite{vojta2} for DS in a homogeneous field. 
The main difference is that for random-field systems
we have to distinguish between
the local damage $d(\varphi)$ which determines  the healing
probability [first term in (\ref{eq:dalin})] and
the average damage $D$ which determines the damaging probability
of a site [second term in (\ref{eq:dalin})].
In a homogeneous system local and average damage are identical.
Consequently, replacing $d(\varphi)$ by $D$ and $m(\varphi)$ 
by $m$ in (\ref{eq:dalin}) 
exactly gives the corresponding equation for the homogeneous system.

To proceed further we have to specify the random field distribution 
$\rho(\varphi)$. As an example  we will discuss the bimodal distribution
\begin{equation}
\rho(\varphi) = \frac 1 2 [\delta(\varphi-\varphi_0) + \delta(\varphi+\varphi_0)]
   \quad (\varphi_0 >0).
\label{distribution}
\end{equation}

\begin{figure}
  \epsfxsize=\figuresize
  \centerline{\epsffile{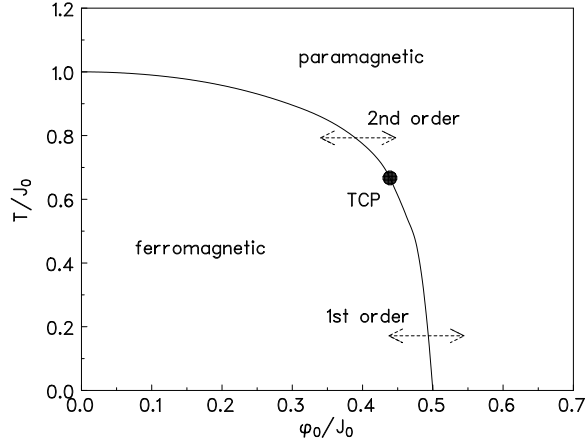}}
  \caption{Thermodynamic phase diagram of the mean-field Ising
              model with bimodal random field. TCP denotes the 
              tricritical point.}
  \label{fig:td}
\end{figure}

The thermodynamics of the mean-field Ising model with a bimodal
random field has been investigated in detail almost 20 years ago 
\cite{aharony}. The equation of state takes the form
\begin{equation}
m= \frac 1 2 \left[ \tanh\left(\frac{J_0 m +\varphi_0} T \right) +
                     \tanh\left(\frac{J_0 m -\varphi_0} T \right) \right].
\label{eq:eqstates}
\end{equation}
The resulting phase diagram is summarized in Fig. \ref{fig:td}.
There is a tricritical point at $T_{\rm TCP} =2J_0/3$ and 
$\varphi_{\rm TCP} \approx 0.439 J_0$.
For $T>T_{\rm TCP}$ the ferromagnetic
phase transition is of second order, for $T<T_{\rm TCP}$ it is of first
order.

We now turn to our results on DS in this model. 
Using the notations $d_{\pm}=d(\pm \varphi_0)$, ${\bf d} =(d_+,d_-)$
and $m_{\pm}=m(\pm \varphi_0)$
the equation
of motion (\ref{eq:dalin}) can be written as
\begin{equation}
\frac d {dt} {\bf d} = {\bf  A} \cdot {\bf d}.
\end{equation}
The dynamical matrix is given by
\begin{equation}
{\bf A} =  \left[
              \begin{array}{cc}
               -|m_+| + (1-m_+^2) \frac {J_0} {2T} & (1-m_+^2) \frac {J_0} {2T} \\
                (1-m_-^2) \frac {J_0} {2T} & -|m_-| + (1-m_-^2) \frac {J_0} {2T} 
              \end{array}
              \right]~.
\end{equation}
The question whether the damage spreads or heals can be answered by means
of the eigenvalues of {\bf A}. If both eigenvalues are negative the damage heals,
if at least one of them is positive the damage spreads. 

\begin{figure}
  \epsfxsize=\figuresize
  \centerline{\epsffile{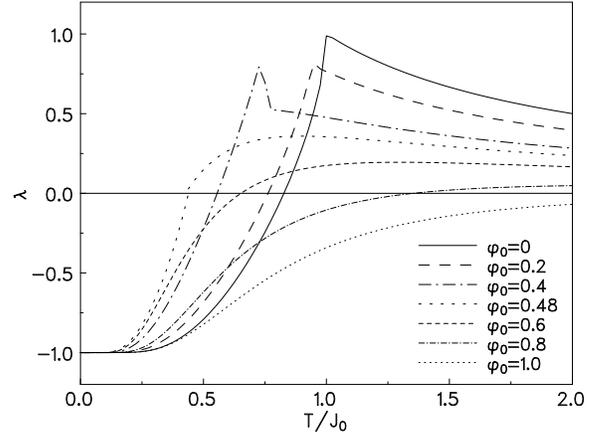}}
  \caption{Lyapunov exponents of the mean-field Glauber
  Ising model with bimodal random field. The peak in the curve for $\varphi_0 =0.4$
  corresponds to the $T$ where $m_-$ vanishes.}
  \label{fig:lambda}
\end{figure}

In the paramagnetic phase we have $|m_+| = |m_-| = \tanh(\varphi_0/T)$.
The eigenvalues of {\bf A} are given by $\lambda_1=-m_+ +(1-m_+^2) J_0/T$
and $\lambda_2= -m_+$. The corresponding eigenmodes are the average
damage $D$ and the damage difference $d_+ - d_-$, respectively. Consequently,
in the paramagnetic phase the Lyapunov exponent which is given by the largest 
eigenvalue of {\bf A} reads
\begin{equation}
\lambda = - \tanh(\varphi_0/T) + [1- \tanh^2(\varphi_0/T)] \, J_0/T.
\label{eq:lambdapara}
\end{equation}
Its dependence on temperature and random field strength is visualized
in Fig. \ref{fig:lambda}.
In the paramagnetic phase the Lyapunov exponent decreases with increasing
random field $\varphi_0$ and therefore the spreading temperature $T_s$ 
which is defined by $\lambda=0$ increases.
This can be easily understood from the fact that a random field favors a
particular spin direction at each site. Since this direction is the same for the two
replicas the corresponding spins in the two replicas tend to be parallel 
which reduces the damage. For random field strength $\varphi_0 >J_0$
the Lyapunov exponent remains negative for all temperatures and thus the
damage never spreads. Asymptotically for $\varphi_0 \to J_0$ we obtain
\begin{equation}
T_s^2 = \frac 2 3 J_0^2 \frac 1 {1-\varphi_0/J_0}.
\end{equation}
We note that the critical $\varphi_0$ which completely
suppresses DS has the same value as the corresponding critical homogeneous field
\cite{vojta2} although the functional dependence of $T_s$ on the field is different.

In the ferromagnetic phase $|m_+|$ and $|m_-|$ are different. In this case
the eigenvalues of {\bf A} are given by
\begin{eqnarray}
\lambda_{1,2} &=& \frac 1 2 \left[-|m_+|-|m_-| + \frac {J_0} {2T} (1-m_+^2 + 1 - m_-^2) \right] 
     \nonumber \\
     &\pm& \left[ \frac 1 4 (|m_+| -|m_-|)^2 + \frac {J_0^2} {16T^2} (1-m_+^2 +1 - m_-^2)^2 \right.
     \nonumber \\
     &&\left. + \frac {J_0} {4T} (|m_+| -|m_-|)(m_+^2-m_-^2)  \right]^{1/2}.
 \label{eq:lambdaferro}
\end{eqnarray}
In order to calculate the Lyapunov exponent we first determine the 
average magnetization $m$ as a function of $\varphi_0$ and $T$
from the equation of state (\ref{eq:eqstates}). We then calculate  
$m_+$ and $m_-$ and insert them into (\ref{eq:lambdaferro}).
The resulting Lyapunov exponents are presented in Fig. \ref{fig:lambda}. 
In contrast to the paramagnetic phase the spreading temperature {\em decreases}
with increasing random field strength.
At a first glance this seems to contradict  the argument
given above, viz. that a random field favors a particular spin direction and thus
reduces the damage. However, the random field also influences 
DS via a reduction of the magnetization since the Lyapunov
exponent (\ref{eq:lambdaferro}) is determined by the local magnetizations.
In the ferromagnetic phase this effect is stronger than that of the
preferred orientation discussed above and 
thus $T_s$ is reduced.

\begin{figure}
  \epsfxsize=\figuresize
  \centerline{\epsffile{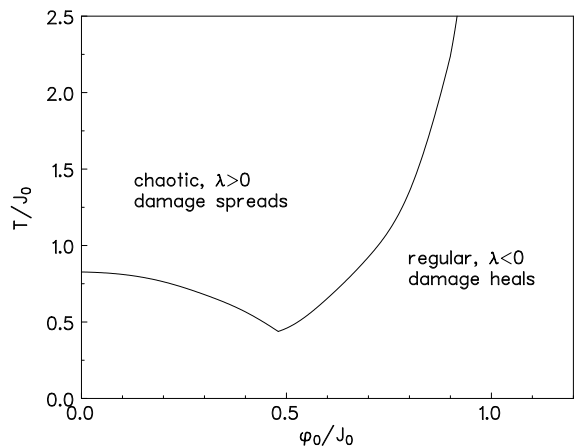}}
  \caption{Damage spreading phase diagram of the mean-field Glauber
  Ising model with bimodal random field}
  \label{fig:pdds}
\end{figure}

By means of (\ref{eq:lambdapara}) and (\ref{eq:lambdaferro}) we have determined 
the spreading temperature
as a function of the random field strength. The resulting phase diagram
of DS in the mean-field Glauber Ising model with bimodal random field is shown
in Fig. \ref{fig:pdds}.
The minimum spreading temperature $T_{s,min}\approx 0.438$ 
is obtained when the spreading
transition coincides with the ferromagnetic phase transition
which occurs for $\varphi_0 \approx 0.480$ (see Fig. \ref{fig:lambda}).

To summarize, we investigated the influence of a quenched random
field on DS in kinetic Ising models. We 
generalized the master equation approach \cite{vojta1,vojta2} to random
field systems and derived an effective field theory for DS. 
As an example we studied the mean-field Glauber Ising model with
bimodal random field.  We found that the random field supports
the spreading of damage in the ferromagnetic phase but hinders
it in the paramagnetic phase. For strong enough field
the damage never spreads. 

In the concluding paragraph we
discuss other random field distributions and
compare our results to the numerical simulation \cite{wappler}.
The influence of the particular form of the random field distribution
on DS can be discussed qualitatively by means of (\ref{eq:dalin}).
This equation shows that the healing probability is proportional
to the local magnetization. This means that the damage on
sites with local magnetization zero cannot heal. 
Consequently, DS will be qualitatively different in 
systems with a continuous random field distribution 
since even for very strong random fields there will be
sites with vanishing local magnetization. Thus
damage will spread on a subset of sites with low enough
random field. However, with $T \to 0$ the 
measure of this subset goes to zero. A detailed 
investigation of this case will be published elsewhere
\cite{vojta3}.
These results also help to understand the numerical 
simulation \cite{wappler} which was carried out for a box
distribution. It shows a decrease of the
spreading temperature with increasing random field although
the stationary value of $D$ is reduced at high temperatures.
This is consistent with a reduction of
$T_s$ due to a suppression of the local magnetization and spreading
on a subset of sites at low temperatures. However, a direct 
comparison with the mean-field theory is not possible since in the 
simulations a two-dimensional system was used which 
-- due to fluctuations -- does 
not have an ordered phase for any finite random field. 
Finally, we discuss possible extensions of this work. 
Besides a systematic investigation of different random
field distributions the damage equation of motion should
be solved beyond first order in the damage. This will 
permit the determination of the stationary damage values
and the investigation of the critical behavior at the spreading
transition. Some studies along these lines are in progress
\cite{vojta3}.

This work was supported in part by the DFG under grant
numbers Vo659/1-1 and SFB393 and by the NSF under grant number 
DMR-95-10185.


\end{document}